\documentstyle[twocolumn,pra,aps,epsfig]{revtex}
\begin{document}
\flushbottom
\draft
\title{Dark states of dressed Bose-Einstein condensates}
\author{E. S. Lee, C. Geckeler, J. Heurich, A. Gupta, Kit-Iu Cheong,
S. Secrest and P. Meystre}
\address{Optical Sciences Center, University of Arizona, Tucson, AZ 85721
\\ \medskip}\author{\small\parbox{14.2cm}{\small \hspace*{3mm}
We combine the ideas of dressed Bose-Einstein condensates, where an intracavity
optical field allows one to design coupled, multicomponent condensates,
and of dark states of quantum systems, to generate a full quantum entanglement
between two matter waves and two optical waves. While the
matter waves are macroscopically populated, the two optical modes share a single
photon. As such, this system offers a way to influence the behaviour of
a macroscopic quantum system via a microscopic ``knob''.
\\[3pt]PACS numbers: 03.75.-b, 03.75.Fi, 42.50.-p}}
\address{}\maketitle
\maketitle
\narrowtext
\section{Introduction}
The recent experimental realization of multicomponent Bose-Einstein condensates
\cite{MyaBurGhr97,StaAndChi98}
has opened up new directions of research, including the study of the 
miscibility and
stability of quantum fluids \cite{SteInoSta98}, 
the nonlinear dynamics of component separation \cite{Tim98},
the generation of spatial patterns \cite{GolMey97,PuBig981,PuBig982}, 
such as e.g.~the antisymmetric collective
mode of two-component condensates \cite{WilWalCoo99}, 
the generation of ferromagnetic and
antiferromegnetic states
\cite{SteInoSta98,OhmMac98,Ho98,LawPuBig98}, etc. 
In particular, the interaction of condensates
with light leads to fascinating effects including the extreme slowing down of
the speed of light\cite{HauHarDut99}, 
matter-wave four-wave mixing \cite{GolPlaMey95,GolPlaMey96,DenHagWen99}, 
and the superradiant scattering of light and atoms 
\cite{MooMey98,MooMey99,InoChiSta99}.

So far, multicomponent condensates have been realized by several 
different methods:
the first one, achieved in $^{87}$Rb \cite{MyaBurGhr97}, 
relies on a fortuitous coincidence
of the scattering lengths of two Zeeman sublevels; the second one, in
$^{23}$Na \cite{StaAndChi98},
uses optical dipole traps to achieve the trapping of the three magnetic
sublevels of the $F=1$ hyperfine ground state. More recently, multicomponent
condensates have also been achieved in nonlinear atom optics experiments
\cite{DenHagWen99}
where the distinction between components is via their center-of-mass
motion rather than their internal state
\cite{StaAndChi98,SteInoSta98,OhmMac98,Ho98,LawPuBig98,GolMey991,GolMey992}. 
In a recent paper
\cite{GolWriMey98} we suggested
yet another approach to the creation of multicomponent condensates, 
relying neither
on distinct electronic nor on center-of-mass levels, but rather on the
dressing of a scalar condensate by the photons of a high-$Q$ optical resonator.

The goal of the present paper is to further examine this idea
in light of the recent work using dark states to achieve coherent quantum
dynamics in condensates and in particular to reduce the speed of light
\cite{HauHarDut99}. In that case, the dark states consist of a coherent
superposition of two hyperfine ground states $|F=1, M_F = -1\rangle$ and
$|F=2, M_F=-2\rangle$ of Sodium, but
the center-of-mass motion of the atoms and the optical fields are treated
classically. In contrast, the present study treats both the optical fields
and the atomic center-of-mass motion quantum-mechanically. Section II
introduces our model and discusses the dark states of a single isolated atom.
This result is extended to the case of dressed condensates in section III.
Section IV considers the effect of ground-state collisions on these states
and evaluates their resulting lifetime. Finally, section V is a summary
and conclusion.

\section{Single-atom dark states}
The experimental situation we have in mind consists of an atomic
sample trapped inside a high-$Q$ optical ring resonator, where the atoms
can interact with two counterpropagating light fields. In addition, they
are subject to spontaneous emission resulting from their interaction
with the continuum of modes of the electromagnetic field. This system
is described by the Hamiltonian
\begin{equation}
H = H_A + H_C + H_R + H_{AC} + H_{AR}
\end{equation}
where the atomic Hamiltonian
\begin{equation}
H_A = \frac{p^2}{2m} + \hbar \omega_0 |e\rangle \langle e| ,
\label{ha}
\end{equation}
contains both the kinetic energy of the atom of momentum ${\bf p}$ and
mass $m$ and its internal energy $\hbar \omega_0$. Note that the form (\ref{ha})
of the atomic Hamiltonian takes the ground states energy to be equal to zero
without loss of generality.

The present section deals with a single atom, which can be taken to be
either a two-level atom or a three-level atom with degenerate ground states
$|g_+\rangle$ and $|g_-\rangle$. However, in section III, which deals with
condensates, it will be necessary to consider a true $\Lambda$-type 
three-level system \cite{MeySar98}. 
Hence, we consider this latter situation from the very beginning. We note
however at the onset that the use of a three-level system does {\em not\/}
imply that we need to consider three different electronic levels: the full
description of the atom requires to specify both its internal and its
center-of-mass state, hence it is possible e.g. to construct a three-level
system consisting of an excited electronic state and a single ground
electronic state, but with two different momenta corresponding, for instance,
to the atomic wave function propagating in opposite directions. For
now, however, we assume for simplicity three electronic levels, since the
extension to other situations is straightforward. We will return to this
point following Eq.~(\ref{scheq}).

We assume that the atom interacts predominantly with two cavity
modes only, further using the dipole approximation and selection rules such that
each of the ground to excited state transitions is driven by just one of
these modes. We then have
\begin{equation}
H_C = \sum_{\mu=\pm} \hbar \omega_\mu {\hat a}_\mu^\dagger {\hat a}_\mu ,
\end{equation}
with $[{\hat a}_\mu, {\hat a}^\dagger_{\mu'}] = \delta_{\mu, \mu'}$, and the
atom-cavity field interaction takes the form
\begin{equation}
H_{AC} = \sum_{\mu=\pm}\hbar {\cal R}_\mu e^{i{\bf k}_\mu \cdot {\bf r}}
|e\rangle \langle g_\mu| \hat a_\mu + H.c.
\label{hac}
\end{equation}
Here, ${\bf r}$ is the center-of-mass location of the atom, with
$[x_i, p_j] = i\hbar \delta_{ij}$ and ${\cal R}_\mu$ is the strength of the dipole
transition between $|g_\mu\rangle$ and $|e\rangle$. Finally, the atom is also
dipole-coupled to the continuum of electromagnetic modes, described by the
Hamiltonian
\begin{equation}
H_R = \sum_i \hbar \omega_i {\hat a}_i^\dagger {\hat a}_i ,
\end{equation}
with $[{\hat a}_i, {\hat a}_j^\dagger] = \delta_{ij}$, and which is responsible 
for spontaneous emission with
\begin{equation}
H_{AR} = \sum_{\mu,i} \hbar {\cal R}_i e^{i{\bf k}_i \cdot {\bf r}} 
|e\rangle \langle g_\mu| \hat a_i + H.c.
\end{equation}
However, we will not need to consider this part of the interaction explicitly
in the present paper.

We consider the situation where the atom is initially so cold that the
momentum width of its center-of-mass wave function is much narrower than
the photon momentum $\hbar k_0$, where $k_0 = \omega_0/c$. This is achieved in
practice by cooling the atom to sub-recoil temperatures. In this case,
the center-of-mass wave function can be treated to an excellent
approximation as a discrete sum of plane waves separated by integer
numbers of photon recoil momenta $\hbar {\bf k}_\mu$. We further assume that
the cavity initially contains just one quantum, which may be in either one of
its modes, and that the atom is initially in the corresponding one of the 
two states of the ground electronic manifold, which interacts with the 
cavity photon. The coherent dynamics described by the interaction
Hamiltonian (\ref{hac}) preserves the number of excitations in the system, but
the quantum initially in the electromagnetic field can be transferred back and
forth between the atom and the cavity modes. In contrast, spontaneous
emission irreversibly couples the one-quantum manifold to the zero-quantum state,
with the atom in its ground state manifold and the cavity in a vacuum.
Hence, for the initial condition at hand we need only consider the one- and
zero-quantum manifolds of states of the atom-cavity system.

Consider first the situation in the absence of spontaneous emission: Because
of the electric dipole selection rules, transitions between
the state $|g_\mu\rangle$ and $|e\rangle$ can be achieved only by absorbing
or emitting a quantum from or into the ``$\mu$''-mode:
\begin{eqnarray}
&& |(g_-, {\bf Q}-{\bf k}_-); 1_-;0_+\rangle
\leftrightarrow |(e, {\bf Q}); 0_-; 0_+\rangle,\\
&& |(g_+, {\bf Q}-{\bf k}_+); 0_-;1_+\rangle
\leftrightarrow |(e, {\bf Q}); 0_-; 0_+\rangle.
\end{eqnarray}
Here, ${\bf Q}$ is the momentum of the atom in its excited state, and
the notation $|(g_-, {\bf Q}) \rangle$, for example, means ``atom in
electronic ground state $|g_-\rangle$ with momentum ${\bf Q}$''. Note that the
photon recoil associated with the dipole interaction (\ref{hac}) has been
explicitly taken into account in these relations, which show that the dipole
interaction between the atom and the cavity modes only couples states within
closed manifolds ${\cal F}_{\bf Q} = \{|\psi_-({\bf Q}) \rangle,
|\psi_+({\bf Q}) \rangle, |\psi_e({\bf Q}) \rangle \}$, with
\begin{eqnarray}
|\psi_-({\bf Q}) \rangle &=& |(g_-, {\bf Q}-{\bf k}_-); 1_-;0_+\rangle \nonumber \\
|\psi_+({\bf Q}) \rangle &=& |(g_+, {\bf Q}-{\bf k}_+); 0_-;1_+\rangle \label{mf}\\
|\psi_e({\bf Q}) \rangle &=& |(e, {\bf Q}); 0_-; 0_+ \rangle. \nonumber
\end{eqnarray}
Within one such manifold, the general state of the system is of the form
\begin{eqnarray}
|\psi({\bf Q}, t) \rangle &=& c_-(t) |\psi_-({\bf Q}, t)\rangle +
\nonumber \\
&& c_+(t) |\psi_+({\bf Q}, t)\rangle +  c_e(t) |\psi_e({\bf Q}, t)\rangle ,
\end{eqnarray}
where the probability amplitudes satisfy the equations of motion
 \begin{eqnarray}
i\hbar \frac{dc_\mu}{dt} &=&\left [ \frac{\hbar^2|{\bf Q} - {\bf k}_\mu|^2}{2m}
+ \hbar \omega_\mu \right ] c_\mu + \hbar {\cal R}_\mu^* c_e ,\nonumber \\
i\hbar \frac{dc_e}{dt} &=& \frac{\hbar^2 Q^2}{2m} c_e
+\sum_{\mu = \pm}  \hbar {\cal R}_\mu c_\mu .
\label{scheq}
\end{eqnarray}

It is useful at this point to return to our earlier discussion of the
two- versus three-level atomic system: It is now quite apparent that since
the ground-state manifold is an electro-translational state, characterized
not just by its electronic state, but also by its center-of-mass quantum
numbers, it could in fact perfectly well correspond to an atom in the
same electronic state, but with different and distinguishable states of
motion. Of course, the dipole selection rules used to express the
interaction Hamiltonian (\ref{hac}) are no longer justified in that case.
However, that same form of interaction, where each field mode interacts with
just one of the atomic transitions in the $\Lambda$-system, can still be
achieved, taking into account the different momenta of the two 
electro-translational ground states instead of the dipole selection rules.
For instance, in the case of counterpropagating cavity modes one can detune
the cavity from the transition frequency of the atom at rest, so that
the Doppler shift associated with its motion brings just one of the states
into resonance with one of the field modes, while the other is shifted
further away from resonance, very much like in Doppler cooling geometries
\cite{Coh92}.
This requires of course that the natural linewidth of the atomic transition
be less than a recoil energy, a condition that can be achieved by using a
long-lived upper electronic state.

We note that the use of an electro-translational ground-state manifold
involving just one electronic state is quite interesting in that it allows one
to create a multicomponent condensate simply by dressing a scalar condensate with a
cavity field as discussed in Ref.~\cite{GolWriMey98}.

In the following we continue to label the electronic ground state(s) with the
symbol $|g_\mu\rangle$, keeping in mind that the single electronic ground state
situation can then readily be obtained by setting $|g_+\rangle = |g_-\rangle$.
None of the remaining algebra is changed then.

So far, we have ignored spontaneous emission. Its effect is to induce
transitions between the one-quantum and the zero-quantum manifold of states,
while imposing a random photon recoil $\hbar {\bf q}$ on the atom. Clearly, 
the realization of a dark state requires that this
effect be eliminated. This can be achieved provided that the probability
amplitude $c_e({\bf Q})$ remains equal to zero for all times. With 
Eqs.~(\ref{scheq}), this implies
\begin{equation}
i\hbar \frac{dc_\mu}{dt} =\left [ \frac{\hbar^2|{\bf Q} - {\bf k}_\mu|^2}{2m}
 + \hbar \omega_\mu \right ] c_\mu ,
\label{eqdark}
\end{equation}
and
\begin{equation}
\sum_{\mu = \pm}  \hbar {\cal R}_\mu c_\mu = 0,
\label{condark}
\end{equation}
Equations (\ref{eqdark}) are readily solved to give
\begin{equation}
c_\mu(t) = c_\mu(0) e^{-i\Omega_\mu t} ,
\end{equation}
with
\begin{equation}
\Omega_\mu =   \frac{\hbar|{\bf Q} - {\bf k}_\mu|^2}{2m} + \omega_\mu .
\end{equation}
However, the additional equation (\ref{condark}) requires that the
probability amplitudes $c_-$ and $c_+$ have a constant phase relation, so that
\begin{equation}
\Omega_+ = \Omega_- \equiv \Omega ,
\end{equation}
which is nothing but a statement of energy-momentum conservation in the Raman-like
transitions between the two ground electronic states. This condition may be
reexpressed as
\begin{equation}
\left [\frac{\hbar k_+}{2m} + u_+ \right ] k_+ =
\left [\frac{\hbar k_-}{2m} + u_- \right ] k_-
\label{cond}
\end{equation}
where
\begin{equation}
u_\mu = c + \hbar \frac{{\bf Q}\cdot {\bf k_\mu}}{m k_\mu}
\label{epcons}
\end{equation}
and the second term on the right-hand side is the component of the atomic recoil
velocity $\hbar {\bf k}_\mu$ along ${\bf Q}$. Typical recoil velocities are
of the order of a few cm/sec, hence $u_\mu \simeq c$, and Eq.~(\ref{cond})
implies that
\begin{equation}
k_+ \simeq k_-
\end{equation}
the equality being exact for $Q = 0$, in which case the corrections
due to the recoil momentum cancel and
\begin{equation}
k_+= k_- \equiv k.
\end{equation}
In that case, the dark state $|\psi_{\mathrm dark}\rangle$ of the system is simply
\begin{eqnarray}
&&|\psi_{\mathrm dark} (Q=0,t)\rangle = \frac{1}{\sqrt{|{\cal R}_+|^2
+ |{\cal R}_-|^2}}\nonumber \\
&\times& {\big [}{\cal R}_-|\psi_+(Q=0)\rangle - {\cal R}_+|\psi_-(Q=0) 
\rangle {\big ]} e^{-i\Omega t} ,
\end{eqnarray}
where $\Omega = \hbar k^2/2m + ck$.

\section{Dressed condensate dark states}

With the single-atom results in mind, we now turn to the situation of
a condensate dressed by the cavity field modes. We assume that the atoms
forming the condensate have the same $\Lambda$-like internal structure as before,
with two degenerate ground states $|g_\mu \rangle$, $\mu = \pm$, which are 
both coupled to a single excited state $|e\rangle$. The condensate is then 
described by a three-component Schr\"odinger field
\begin{equation}
{\hat{\bbox  \psi}}({\bf r})  = \left ( \begin{array}{c}
{\hat \psi}_-({\bf r})\\{\hat \psi}_+({\bf r})\\ {\hat \psi}_e({\bf r})
\end{array}
\right )
\end{equation}
with $[{\hat \psi}_i({\bf r}), {\hat \psi}_j^\dagger({\bf r}')] =
\delta_{ij} \delta({\bf r} - {\bf r}')$ and $i, j = \pm, e$. Because of the
central role photon recoil plays in this problem, it is convenient to
expand these components in terms of plane waves as
\begin{equation}
{\hat \psi}_i ({\bf r}) = \frac{1}{\sqrt{V}} \sum_{\bf q}
e^{i{\bf q}\cdot{\bf r}} {\hat c}_{i,{\bf q}}
\end{equation}
with
\begin{equation}
[{\hat c}_{i,{\bf q}}, {\hat c}^\dagger_{j, {\bf q}'}] = \delta_{i j}
\delta({\bf q} -{\bf q}') ,
\end{equation}
in terms of which the second-quantized version ${\cal H}_0$ of the Hamiltonian
$H_0 = H_A + H_C + H_{AC}$ is
\begin{eqnarray}
&&{\cal H}_0 =
\sum_{i, {\bf q}} \frac{\hbar^2 q^2}{2m} {\hat c}^\dagger_{i, {\bf q}}
{\hat c}_{i, {\bf q}} + \sum_{\bf q} \hbar \omega_0
{\hat c}^\dagger _{e, {\bf q}}{\hat c}_{e, {\bf q}} \nonumber \\
&&+ \sum_\mu \hbar \omega_\mu {\hat a}^\dagger_\mu {\hat a}_\mu
+ \sum_{\mu, {\bf q}} \left (\hbar {\cal R}_{\mu} {\hat a}_\mu
{\hat c}^\dagger_{e, {\bf q}}
{\hat c}_{\mu, {\bf q}-{\bf k}_\mu} + H.c.\right )
\end{eqnarray}
where $i = \pm, e$ and $\mu = \pm$.

We consider a condensate consisting of $N$ atoms, and assume as before that
there is at most one quantum of excitation in the condensate-cavity system.
It is not difficult to generalize the single-atom discussion to this case.
Instead of the family of closed manifolds ${\cal F}_{\bf Q}$ used
for the single-atom case we now need to consider the extended set
of manifolds ${\cal F}_{\bf Q}(N_-, N_+)$. Indeed, it is easily shown that
the manifold ${\cal F}_{\bf Q}(N_-, N_+) = \{|\psi_-({\bf Q}, N_-, N_+) \rangle,
|\psi_+({\bf Q}, N_-, N_+) \rangle, |\psi_e({\bf Q}, N_-, N_+) \rangle \}$ forms a closed set
of states for the condensate-cavity evolution we get from the atom-cavity field interaction described in ${\cal H}_0$,
where
\begin{eqnarray*}
&&|\psi_-({\bf Q}, N_-, N_+) \rangle = \\
&&|(N_-,{\bf Q} - {\bf k}_-);(N_+-1,{\bf Q} - {\bf k}_+);
(0_e,{\bf Q});1_-;0_+\rangle \\
&&|\psi_+({\bf Q}, N_-, N_+) \rangle = \\
&&|(N_--1,{\bf Q} - {\bf k}_-);( N_+,{\bf Q} - {\bf k}_+ );
(0_e, {\bf Q});0_-;1_+\rangle \\
&&|\psi_e({\bf Q}, N_-, N_+) \rangle = \\
&&|(N_--1,{\bf Q} - {\bf k}_-);(N_+-1,{\bf Q} - {\bf k}_+);
(1_e, {\bf Q});0_-;0_+\rangle
\end{eqnarray*}
and $N_+ + N_- =N + 1.$

In the following we restrict our discussion to the case ${\bf Q} =0$,
${\bf k}_- = -{\bf k}_+ \equiv {\bf k}$, $k = \omega/c$. The extension to more general situations
is straightforward. For notational clarity we also temporarily drop the variables
${\bf Q}$ as well as $N_-$ and $N_+$ from the definitions of the states, and express the
state of the system as in the single-atom case as
\begin{equation}
|\psi(t) \rangle = c_-(t) |\psi_-\rangle + c_+(t) |\psi_+\rangle +
c_e(t) |\psi_e\rangle .
\end{equation}
The equations of motion for the probability amplitudes $c_i(t)$ are now
\begin{eqnarray}
i\hbar \frac{dc_\mu}{dt} &=&\left [ N\frac{\hbar^2 k^2}{2m}
+ \hbar \omega_\mu \right ] c_\mu + \hbar {\cal R}_\mu^*\sqrt{N_\mu} c_e ,\nonumber \\
i\hbar \frac{dc_e}{dt} &=& (N-1)\frac{\hbar^2 k^2}{2m} c_e
+\sum_{\mu = \pm}  \hbar {\cal R}_\mu \sqrt{N_\mu} c_\mu ,
\label{scheq2}
\end{eqnarray}
where $\mu = \pm$.

Despite the fact that we are now considering an $N$-atom problem, the selection 
rules of the atomic system and the choice of the initial condition result in these
equations having the same form as the single-particle equations (\ref{scheq}),
with two differences: First, the dipole coupling constants ${\cal R}_\mu$ are
now replaced by
\begin{equation}
{\cal R}_\mu \rightarrow {\cal R}_\mu \sqrt{N_\mu} ,
\end{equation}
a result of the collective action (Bose enhancement) from the $N_\mu$ atoms
in the $\mu$'s condensate component. Second, the kinetic energy term now
accounts for all $N$ atoms, the $(N-1)$-factor in the equation for $c_e$
resulting from photon recoil. With these changes, one can readily adapt the
results of section II, and find that the many-particle dark state associated
with a given manifold $\{Q = 0, N_-, N_+\}$ is
\begin{eqnarray}
&&|\psi_{\mathrm dark} (Q=0,N_-, N_+, t)\rangle = 
\frac{1}{\sqrt{N_+|{\cal R}_+|^2 + N_-|{\cal R}_-|^2}}
\nonumber \\
&\times& \left [\sqrt{N_-}{\cal R}_-|\psi_+(Q=0, N_-, N_+)\rangle 
\right . \nonumber \\
&-& \left .\sqrt{N_+} {\cal R}_+|\psi_-(Q=0, N_-, N_+) \rangle \right ] 
e^{-i\Omega_N t},
\label{ds}
\end{eqnarray}
where
\begin{equation}
\Omega_N = N \frac{\hbar k^2}{2m} + ck.
\label{en}
\end{equation}
Hence, dark states of the multicomponent condensate can exist for any value of
their relative populations. However, there is an important distinction
between the present case and the single-atom situation, because of the effect
of ground state collisions, that we have ignored so far. For ultracold
atoms, they can be described in the
shapeless approximation by a local
two-body interaction, that is, a very broad potential in momentum space.
Ground-state collisions therefore result in velocity changes in the atoms,
and hence in the atoms escaping from the dark states (\ref{ds}). The next
section evaluates the lifetime associated with these collisions.

\section{The effect of ground-state collisions}
In the shapeless, $s$-wave scattering approximation, the ground-state
collisions are described by the Hamiltonian
\begin{equation}
{\cal V} = \frac{4\pi \hbar^2 a}{mV} \sum_{\mu, \mu^\prime = \pm} \; \sum_{{\bf q}, {\bf q}',
{\bbox \kappa}}
{\hat c}^\dagger_{\mu, {\bf q}-{\bbox \kappa}}
{\hat c}^\dagger_{\mu', {\bf q}'+{\bbox \kappa}}
{\hat c}_{\mu', {\bf q}'}
{\hat c}_{\mu, {\bf q}},
\end{equation}
where $a$ is the scattering length of the collisions, assumed to be the
same for both ground states.

The $\kappa=0$ contribution to ${\cal V}$ corresponds to velocity-conserving
collisions, which do not lead to an escape of the condensates from the
dark state. Hence, it is useful to treat it separately. Reexpressing it
as
\begin{eqnarray}
{\cal V}_0 &=& \frac{4\pi \hbar^2 a}{mV} \sum_{\mu, \mu^\prime = \pm} \; \sum_{{\bf q},{\bf q}'}
{\hat c}^\dagger_{\mu, {\bf q}}{\hat c}^\dagger_{\mu', {\bf q}'}
{\hat c}_{\mu', {\bf q}'}{\hat c}_{\mu, {\bf q}}  \nonumber \\
&=& \frac{4\pi \hbar^2 a}{mV} \sum_{\mu, \mu^\prime = \pm} \; \sum_{{\bf q},{\bf q}'}
{\hat c}^\dagger_{\mu, {\bf q}}[{\hat c}_{\mu, {\bf q}}{\hat c}^\dagger_{\mu', {\bf q}'}
- \delta_{\mu, \mu'}\delta_{{\bf q}, {\bf q}'}] {\hat c}_{\mu', {\bf q}'}
  \nonumber \\
&=& \frac{4\pi \hbar^2 a}{mV} ({\hat N}^2_g  - {\hat N}_g) ,
\end{eqnarray}
where ${\hat N}_g$ is the number operator for all ground-state atoms, we
observe that the states $|\psi_i({\bf Q}, N_-, N_+)\rangle$ are eigenstates
of that operator, specifically,
\begin{eqnarray}
{\cal V}_0 |\psi_-\rangle &=& \frac{4 \pi \hbar^2 a}{m V} (N^2-N) |\psi_-\rangle  , \nonumber \\
{\cal V}_0 |\psi_+\rangle &=& \frac{4 \pi \hbar^2 a}{m V} (N^2-N) |\psi_+\rangle  , \nonumber \\
{\cal V}_0 |\psi_e\rangle &=& \frac{4 \pi \hbar^2 a}{m V}(N^2-3N+1) |\psi_e\rangle  .
\end{eqnarray}
Hence, its effect is merely to change the eigenenergy $\hbar \Omega_N$ of the
dark state, see Eq.~(\ref{en}), to
\begin{equation}
\hbar \Omega_N  = N\frac{\hbar^2 k^2}{2m} + \hbar ck + \frac{4\pi\hbar^2 a}{mV} (N^2 -N).
\label{encol}
\end{equation}

The remaining collisions, described by ${\cal V\,}^\prime$, couple the atoms in the dark state of the
condensate to a continuum of momentum states, which is essentially flat
in the shapeless approximation. These momentum states may be thought of
as a bath, and the decay of the dark state into that reservoir can be evaluated
via Fermi's Golden Rule. Hence, we need to keep only those states coupled to
the dark state $|\psi_{\mathrm dark}\rangle$  by ${\cal V\,}^\prime$ that satisfy energy
conservation. To proceed it is convenient to separate ${\cal V\,}^\prime$ into
the contributions involving atoms in the same and in different electronic
ground states. In the former case, conservation of energy between the initial
and final state requires in general that
\begin{equation}
q^2 + q'^2 = |{\bf q} - {\bbox \kappa}|^2 + |{\bf q}' + {\bbox \kappa}|^2 .
\label{consen}
\end{equation}
But atoms in the dark state $|\psi_{\mathrm dark}\rangle$ and in a given ground
electronic state all have the same momentum, so that
${\bf q} = {\bf q}'$, and Eq.~(\ref{consen}) reduces simply to
$\kappa=0$: that is, the only energy-conserving collisions between atoms in
the same ground state are velocity-preserving. But these collisions are
already accounted for in the modified eigenenergy (\ref{encol}).
Hence, the lifetime of the condensate dark states is determined solely by
collisions involving atoms in different ground states, $\mu \ne \mu'$.

For these collisions, the conservation of energy condition takes the form
\begin{equation}
q_-^2 + q_+^2 =
|{\bf q}_- - {\bbox \kappa}|^2 + |{\bf q}_+ + {\bbox \kappa}|^2 ,
\label{consen2}
\end{equation}
where ${\bf q}_\mu$ is the momentum of dark state atoms in the ground
electronic state $|g_\mu\rangle$.

In case the two cavity modes are counterpropagating, ${\bf k}_+ = -{\bf k}_-$
and for ${\bf Q} = 0$, Eqs.~(\ref{mf}) show that ${\bf q}_+ = -{\bf q}_- = {\bf k}$,
so that Eq.~(\ref{consen2}) reduces to
\begin{equation}
k^2 = |{\bf k} + {\bbox \kappa}|^2 .
\end{equation}
Energy conservation is therefore satisfied for transferred momenta ${\bbox \kappa}$
on a sphere of radius $k$ and centered at $-{\bf k}$. Noting that each possible state
occupies a volume of $8\pi^3/V$ in momentum space, we have that the density
of states on the energy shell
\begin{equation}
\epsilon \equiv \frac{\hbar^2 k^2}{2m}
\label{epsi}
\end{equation}
is
\begin{equation}
{\cal D}(\epsilon) =
\frac{d{\cal N}(\epsilon)}{d\epsilon} = \frac{1}{2\pi^2 \hbar^3}
(2m)^{3/2} \sqrt{\epsilon} .
\end{equation}
With this result, and in the limit of large condensates $N_+, N_- \gg 1$,
a staightforward Fermi Golden Rule calculation yields the dark state
decay rate
\begin{equation}
\Gamma = 32\pi \frac{\hbar k a^2}{mV} N_+N_- ,
\end{equation}
where we have used Eq.~(\ref{epsi}) to express the energy $\epsilon$ in
terms of the photon recoil momentum $\hbar k$. For tyical condensates
containing of the order of $10^6$ atoms at a density of $10^{12}$ cm$^{-3}$,
a scattering length $a \simeq 10^{-7}$ cm and visible light we find that
$\Gamma \simeq 10^6$ sec$^{-1}$. For such relatively short lifetimes, it is
acceptable to neglect the effects of optical cavity loss as we have done
here, provided that one uses the high-$Q$ cavities familiar e.g.\ from cavity
QED experiments. Since $\Gamma$ scales as the square of the number of
atoms in the condensate, longer lifetimes can easily be achieved, but in
that case, cavity losses are expected to start playing an important role
and to dominate the lifetime of the condensate dark states.

\section{Summary and conclusion}
Dressing scalar condensates with electromagnetic fields allows one to design
coupled, multicomponent condensates that permit to study the dynamics
of coupled matter-wave fields under well-controled conditions. In addition,
coherent control techniques such as electromagnetically induced transparency
allow one to manipulate the optical properties of condensates in dramatic ways,
opening up the way to novel investigations involving e.g.\ the coupling of slow
optical waves and accoustic waves, etc. In this paper, we have combined
these two techniques to realize dark states of dressed condensates. These
states exhibit full quantum mechanical entanglement between the four
``modes'' involved, two matter waves and two optical modes. While the
matter waves are macroscopically populated, the two cavity modes share a single
photon. As such, this system offers a way to influence the behaviour of
a macroscopic quantum system via a microscopic ``knob'' and might be of
use in elucidating fundamental questions in quantum measurement theory
and quantum information processing. In particular, this is precisely the
situation Schr\"odinger had in mind in his cat ``paradox.''

\acknowledgements
We thank E. V. Goldstein for numerous discussions and comments. 
This work was supported in part by the U.S. Office of Naval Research
Contract No. 14-91-J1205, by the U.S. Army Research Office, by the
Joint Services Optics Program and by the National Science Foundation.


\end{document}